\documentclass[%
superscriptaddress,
 amsmath,amssymb,
prf,
onecolumn,
]{revtex4-2}
\usepackage[export]{adjustbox}
\usepackage{graphicx}% Include figure filessn
\usepackage{color}% Include figure files
\usepackage{xcolor}% Include figure files
\usepackage{natbib}
\usepackage{wasysym,amsmath,mathtools}% Include figure files
\usepackage{dcolumn}% Align table columns on decimal point
\usepackage{bm}% bold math
\usepackage{hyperref}% add hypertext capabilities
\definecolor{mygreen}{rgb}{0,0.5,0}
\definecolor{mybrown}{rgb}{0.65,0.16,0.16}

\def\beq {\begin{equation}}
\def\eeq {\end{equation}}
\def\beqa {\begin{eqnarray}}
\def\eeqa {\end{eqnarray}}
\def \bnum {\begin{enumerate}}
\def \enum {\end{enumerate}}
\def\bi {\begin{itemize}}
\def\ei {\end{itemize}}

\def\la {\langle}
\def\ra {\rangle}

\def\mbf {\mathbf}
\def\uu{\mbf{u}^\nu}
\def\ux{\mbf{x}}
\def\ur{\boldsymbol{\ell}}

\def\up{{u^\prime}}
\def\nabv{\mbf{\nabla}}
\def\dru{\delta_\ell u^\nu_\parallel}
\def\druv{\boldsymbol{\delta}_{\mbf{\ell}} {\uu}}

\def\epsn{{\epsilon^\nu}}

\def\epstar{{\epsilon^*}}
\def\up{{u^\prime}}
\def\elo{\ell_0}
\def\sr{S^\parallel_3(\ell)}
\def\ar{A^\parallel_3(\ell)}
\def\epsnrlo{\frac{\epsn}{\la|\uu |^3 \ra}}
\def\epsnrlone{\frac{\epsn}{\la|\uu |^3 \ra}}
\def\epsnrlonenofrac{{\epsn}/{\la|\uu |^3 \ra}}

\def\lu{\frac{1}{\la|\uu |^3\ra}}

\def\xin{\xi_{3,\parallel}}
\def\xstar{\xi_{3,\parallel}}

\def\ellint{\ell_{int}}
\def\epsnrlint{\frac{\epsn \ellint}{\la|\uu |^3 \ra}}
\def\epsnrnofraclint{{\epsn\ellint}/{\la|\uu |^3 \ra}}
\def\xintot{\xi_{3}}

\begin{document}
\title{Dependence of the asymptotic energy dissipation on third-order velocity scaling}
%%%%%%%%%%%%%%%%%%%%%%%%%%
\author {Kartik P. Iyer} 
\affiliation {Department of Physics, Michigan Technological University, Houghton, MI $49931$}
\affiliation {Department of Mechanical Engineering-Engineering Mechanics, Michigan Technological University, Houghton, MI $49931$}
\email{kiyer@mtu.edu}

\begin{abstract}
The asymptotic energy dissipation is connected to the 
third-order scaling of the longitudinal velocity increment magnitude
in three-dimensional turbulence via the Kolmogorov $4/5$ law. It is shown that the third-order longitudinal absolute velocity increment scaling should not exceed unity for anomalous dissipation to occur, that is for non-vanishing average dissipation in the inviscid limit -- also known as the ``zeroth law" of turbulence. 
Conversely, if the third-order longitudinal absolute velocity increment scaling exceeds unity then the average dissipation must asymptotically vanish and the velocity increment field will becomes symmetric at least at the level of its skewness. This work highlights the importance of the third-order absolute velocity increment scaling in assessing the status of the ``zeroth-law" of turbulence.  
\end{abstract}
\maketitle

A surprising phenomenon in three-dimensional incompressible turbulence is that the average energy dissipation does not seem to decay with increasing Reynolds number. To account for this enhanced dissipation, Lars Onsager asserted that if the 
spatial H{\"o}lder exponents of the velocity field are utmost one-third then the average dissipation can be non-zero in the invisicd limit \cite{LO49}.  
{Following a result by Eyink \cite{GL94}, Onsager's theorem was fully proved by Constantin et al.~\cite{CWT}. In particular, it was shown that if the spatial H{\"o}lder exponents exceeded one-third then the average energy dissipation must vanish and energy will be conserved in the inviscid limit \cite{CWT}.
The Besov space formulation of \cite{CWT} also meant that if the third-order exponent of the velocity increment magnitude moment exceeded unity then the average dissipation must vanish asymptotically in the inviscid limit.}
Almost all such theoretical studies on the asymptotic dissipation starting from that of Onsager until now have related the energy dissipation to the scaling properties of the total velocity increment magnitude field \cite{LO49,GL94,CWT,TDGE2019}. However due to practical considerations, it is the projections of the total velocity increments along the separation distances - known as the longitudinal velocity increments, that are routinely measured in experiments and simulations \cite{SA97,ishihara09}. 

\vspace{0.2cm}
The purpose of this short Letter is to connect the average energy dissipation to the third-order scaling exponent of the longitudinal velocity increment magnitude moment. 
This third-order connection between energy dissipation and the longitudinal absolute velocity increment scaling is contrasted to the already known connection between energy dissipation and the total absolute velocity increment scaling \cite{CWT,TDGE2019}.
The implications of this result for the asymmetry of the small-scale velocity field in the inviscid limit are also discussed.

\vspace{0.2cm}
{Consider the three-dimensional divergence-free velocity $\uu \coloneqq \uu(\ux,t)$ in a boundary-less domain such as the $3$-torus ${\mathbb{T}}^3$}. Here $\nu$ is the kinematic viscosity of the fluid,
$\ux$ denotes position and $t$ denotes time.
Define the Reynolds number $Re = \up \elo/\nu$
where,
$\elo $ is a (fixed) large length scale,
$\up = {\la |\uu|^2 \ra}^{1/2}$
is the root-mean-square velocity and
$\la \cdot \ra$ denote space-time averages.
The average energy dissipation rate is 
given by 
\beq
\label{diss.eq}
\epsn \coloneqq \nu \la |\nabv \uu|^2 \ra \ge 0 \;.
\eeq 

Consider the total velocity increment $\druv$ and its longitudinal component $\dru$ between two distinct points separated by vector $\ur \in \mathbb{R}^3$ at 
distance $\ell = |\ur| > 0 \in \mathbb{R}$ at given time $t$,
\beqa
\label{aa.eq}
\druv &\coloneqq& \uu(\ux+\ur,t)-\uu(\ux,t) \;, \\
\label{bb.eq}
\dru  &\coloneqq& \druv \cdot {\ur}/{|\ur|} \;. 
\eeqa
At order three, the {longitudinal velocity increment moment} and the longitudinal absolute velocity increment moment -- also known as the 
longitudinal structure function and the longitudinal absolute structure function respectively are defined for any separation $\ell$ as,
\beq
\label{strucfns.eq}
{
\sr \coloneqq \la (\dru)^3 \ra} \;, \quad \ar \coloneqq \la |\dru|^3 \ra \;.
\eeq
%$\sr \coloneqq |\la (\dru)^3 \ra|$. The
%third-order longitudinal absolute structure function $\ar \coloneqq \la |\dru|^3 \ra$.
We note for later use that at any scale $\ell$ the two structure functions
are related by the triangle inequality as,
%Denote the magnitude of the third-order moment of the longitudinal increment
%and third-order moment of the absolute longitudinal increment $|\dru|$,
\beq
\label{s3.eq}
{\sr \le |\sr|}\le \ar  \;. 
%\;, \quad A_3(r) = \la |\dru|^3 \ra \;,
\eeq

In what follows, we non-dimensionalize all physical quantities using $\elo$, $\up$ and $\elo/\up$ as the relevant length, velocity and time scales respectively. 
%We can then set the dimensionless large length scale 
%$\ell_0 = 1$ and the 
%dimensionless velocity scale $\up = 1$. 
The dimensionless viscosity becomes the inverse Reynolds number $\nu = 1/Re$ and the asymptotic limit $Re \to \infty $ is equivalent to 
$\nu \to 0$. 
%This normalization yields,
{Using the H\"older inequality this normalization gives},
\beq
%\label{l.eq}
 %  0 & < \ell \le   1 \;, \\
 \label{u.eq}
1     =  \up^3   \le   \la |\uu|^3 \ra  \;.
\eeq
%the last inequality stems from the H\"older inequality.   

\vspace{0.2cm}
For $\ell > 0$ we start with the trivial identity
\beq
\label{tr.eq}
\frac{4}{5}\epsn= \frac{\sr}{\ell} + \frac{4}{5}\epsn
- \frac{\sr}{\ell} \;.
\eeq
Dividing both sides of \eqref{tr.eq} by $\la |\uu|^3 \ra > 0$, using the triangle inequality and \eqref{s3.eq} we can bound the left-hand-side of \eqref{tr.eq} as,
\beq
\label{tro.eq}
\frac{4}{5}\epsnrlo \le \frac{\ar}{\ell}\lu + 
\Big |\frac{4}{5}\epsn - \frac{\sr}{\ell} \Big | \lu \;.
\eeq
{Using \eqref{u.eq} we can write \eqref{tro.eq} as},
\beq
\label{trm.eq}
\frac{4}{5}\epsnrlo \le \frac{\ar}{\ell}\lu + 
\Big |\frac{4}{5}\epsn - \frac{\sr}{\ell} \Big | \;.
\eeq
Since it follows from \eqref{bb.eq} that the longitudinal absolute velocity increments cannot exceed the total velocity increment magnitudes at any $\ell$, i.e.~$|\dru| \le |\druv|$, their corresponding third-order moments moments are related as,
%Since the third-order longitudinal absolute velocity structure function is bounded by the total absolute velocity structure function at any $\ell$, we can write,
%assume that $A_3(\ell)$ can be bounded by a power-law,
%%%%%%%%%%%%
%assumes a power-law behavior in the range $\eta \ll \ell \ll 1$ which extends to ever-decreasing $\ell$ as $\eta$ decreases with decreasing $\nu$
%as shown in Fig.~\ref{f1.fig} we can write,
\beq
\label{mink.eq}
\ar \le \la |\druv|^3 \ra \le 8 \la |\uu|^3 \ra  \;,
\eeq
where the last inequality follows from Minkowski's inequality. 
{We now assume the following power-law bound,}
%%%%%%%
\beq
\label{pl.eq}
{
\frac{\ar}{\la |\uu|^3 \ra } \le  8 {\ell}^{\xin} 
\;, 
}
\eeq
where $\xin > 0$ is the third-order longitudinal absolute scaling exponent, $\ar \propto \ell^{\xin}$.
{Our principle assumption here is that bound \eqref{pl.eq} holds uniformly in 
viscosity $\nu \in (0,1]$.
Substituting the upper bound \eqref{pl.eq} into right-hand-side of \eqref{trm.eq} we 
get}
\beq
\label{uboundbefk41.eq}
\frac{4}{5}\epsnrlone \le 8 \ell^{(\xin-1)} + 
\Big |\frac{4}{5}\epsn - \frac{\sr}{\ell} \Big | \;.
\eeq
In order to obtain the asymptotic dissipation
we first send $\nu \to 0$ and then send $\ell \to 0$ in \eqref{uboundbefk41.eq}. 
Since the  
left-hand-side in 
\eqref{uboundbefk41.eq} is $\ell$-independent and $\xin$ is assumed to be $\nu$-independent we get,
\beq
\label{uboundlim1.eq}
{
\frac{4}{5} \limsup_{\nu \to 0} \epsnrlone \le 
8 \lim_{\ell \to 0} 
\ell^{(\xin-1)}
+ 
\lim_{\ell \to 0} \lim_{\nu \to 0}
\Big |\frac{4}{5}\epsn - \frac{\sr}{\ell} \Big | \;.
}
\eeq
Noting that the second term on the right-hand-side in 
\eqref{uboundlim1.eq} vanishes as it is the precise formulation of the Kolmogorov $4/5$ law \cite{K41b,Nie99,GE2003,Bedrossian19,TD22} we get,
\beq
\label{dissbound.eq}
{
\limsup_{\nu \to 0} \epsnrlone \le
10 \lim_{\ell \to 0} 
\ell^{(\xin-1)} \;.
}
\eeq
Denoting the {limit superior} of the 
normalized 
dissipation as follows, 
\beq
\label{asymplong.eq}
{
\limsup_{\nu \to 0}  \epsnrlone \coloneqq \epstar \;,
}
\eeq
we can finally write \eqref{dissbound.eq} as,
\beq
\label{dissboundfinal.eq}
\epstar  \le
10 \lim_{\ell \to 0}
\ell^{(\xstar-1)} \;.
\eeq
{From \eqref{dissboundfinal.eq} it follows that if
$\xstar > 1$ then $\epstar = 0 $ and so the limit of the normalized dissipation exists and is zero}. That is,
\beq
\label{result.eq}
\textrm{If} \;\; \xstar > 1 \;\; \implies \;
\epstar = \lim_{\nu \to 0}  \epsnrlone = 0 \;.
\eeq
In this case the asymptotic normalized dissipation vanishes and energy is conserved in the asymptotic $\nu \to 0$ limit. It follows from \eqref{result.eq} that a necessary (but not sufficient) condition for $\epstar$ to be non-zero is that $\xstar \le 1$. We note that although \eqref{dissboundfinal.eq} clarifies the fate of the asymptotic dissipation $\epstar$, it does not provide a conditional decay rate for dissipation. Such a conditional dissipation decay rate is provided in \cite{TDGE2019} in terms of the third-order total absolute structure function exponent ${\xintot}$, where $\la |\druv|^3 \ra \propto \ell^{{\xintot}}$.  

\vspace{0.2cm}
Furthermore, in isotropic turbulence since the integral scale $\ell_{int}$ is typically defined as \cite{pope}
\beq
\label{lint.eq}
\ell_{int} \coloneqq
\frac{3}{2}\frac{\pi} {{\up}^2}\int_0^\infty \frac{E(\kappa)}{\kappa} d\kappa \le 1 \;,
\eeq
where $\kappa$ is the wavenumber magnitude and
$E(\kappa)$ is the three-dimensional energy spectrum.
%and the 
%fixed $\elo$ is chosen such that $\ell_{int} \le \elo = 1$. % to yield \eqref{lint.eq} . 
If the asymptotic dissipation $\epstar$ vanishes, that is if \eqref{result.eq} holds, then the asymptotic normalized dissipation defined using $\ellint$ must also vanish since \eqref{lint.eq} implies, 
\beq
\label{dissboundlint.eq}
\lim_{\nu \to 0}  \epsnrlint \le \epstar  \;. 
\eeq
The upper bound \eqref{dissboundlint.eq} is especially
useful in Direct Numerical Simulations where the evolution of $\ell_{int}$ is often 
undercut by limited domain sizes $(\sim \elo^3 )$ \cite{Kops}. In such a scenario an examination of $\epsnrlonenofrac$ rather than that of $\epsnrnofraclint$ may be more insightful since if the former vanishes then the latter must also disappear due to \eqref{dissboundlint.eq}. 
%
%\vspace{0.2cm}
%Finally, we note that in principle 
%$\la |\uu|^3 \ra $ can diverge as $\nu \to 0$, i.e.~the third-order velocity magnitude moment can have some non-trivial Reynolds number dependence \cite{Bedrossian19}. However, empirically one can expect this third-order moment $\la |\uu|^3 \ra$ to be a $\nu$-independent constant since the probability density function of $\uu$ is essentially Reynolds-number-independent. Under such an assumption, \eqref{result.eq} can be expected to hold for the non-dimensional asymptotic dissipation since in this case $\lim_{\nu \to 0}\epsilon^\nu \coloneqq \epstar $. 

%Finally we note that since the probability density function of $\uu$ is likely to be at most weakly Reynolds-number-dependent, on empirical grounds one can expect $\la |\uu|^3 \ra \propto  1$.

\section{Discussion}
The asymptotic behavior of turbulent dissipation is not only 
a problem of fundamental importance but it is also relevant to energy considerations in modelling turbulent drag in applications such as aerodynamics and fluid transport in pipelines \cite{GEKRS,vas15,BD19}. Until now 
the asymptotic dissipation $\epstar$ 
has been connected to the third-order total scaling exponent $\xintot$ which is seldom examined in empirical work.
In this work we have shown that
under the assumption of the Kolmogorov $4/5$ law,
the asymptotic dissipation $\epstar$ must vanish
if the third-order longitudinal absolute exponent $\xstar > 1$. The significance of this work is as follows. 

\vspace{0.2cm}
At order three, the longitudinal absolute exponent $\xin$ is larger than the total absolute exponent ${\xintot}$
%where $\la |\druv|^3 \ra \propto \ell^{\xi^\nu_{3,T}} $
%i.e.~$\la |\dru|^3 \ra \le \la |\druv|^3 \ra $ and
i.e.~$\xin \ge {\xintot}$. This is because $\la |\druv|^3 \ra$ includes the transverse velocity difference component which is known to be more intermittent with a smaller associated exponent even at order three \cite{DTS97}.
It then follows from this work that the asymptotic dissipation must vanish even if the total absolute exponent ${\xintot} \le 1$, as long as
the Kolmogorov $4/5$ law is valid and $\xstar > 1$. In this sense this result is a sharper result than that of \cite{CWT,TDGE2019}. 
%It will be useful to quantify the dissipation decay rate in terms of $\xstar$ as has been done in \cite{TDGE2019} 
%however that if 
%$\xi^\nu_{3,T} > 1 $ then \cite{TDGE2019} provides a dissipation decay rate which, it will

\vspace{0.2cm}
Another implication of this work is that for the asymptotic longitudinal velocity difference field. If $\xstar > 1$ and $\epstar = 0$ then it follows from the exact Kolmogorov $4/5$ law that asymptotically $\sr/\ell \to 0$ \cite{Bedrossian19,TD22}. Since the longitudinal velocity increment is known to scale linearly $\sr \propto \ell^1$ in the inertial range 
\cite{krs98,KIKRS20}, it must follow that if $\xstar > 1$ then the third-order longitudinal structure function
$\sr \to 0$ as $\nu \to 0$ due to cancellations in its power-law prefactor. This implies that if $\xstar > 1$ then the velocity increment field will asymptotically become symmetric at least at the level of its skewness.
{In the Lagrangian context, this symmetrization will result in the asymptotic mean-square particle displacement becoming time reversible \cite{TDlag}. 
This space-time recovery of reversibility at least up to order three in the vanishing dissipation case can have important consequences in modelling the asymptotic limit $\nu \to 0$. 
}

\vspace{0.2cm}
In the alternate scenario where $\xstar \le 1$ and $\epstar > 0$,
the small-scale asymmetry will persist at all non-trivial orders. 
{In particular it will follow from the Kolmogorov $4/5$ law that the 
velocity increment field will have non-vanishing (negative) skewness in the asymptotic limit, $\sr <0 $ as $\nu \to 0$. The non-zero asymptotic dissipation will manifest in temporal asymmetry of the kinetic energy of particles in the asymptotic limit $\nu \to 0$, that is different short time particle dispersion in forward and backward time in the Lagrangian setting \cite{TDlag}.}

\vspace{0.2cm}
Finally, a few remarks about the Reynolds number scaling of the asymptotic dissipation from experiments and simulations are in order. 
A majority of the empirical studies with few exceptions have observed a non-trivial independence of the normalized turbulent energy dissipation on the Reynolds number - this phenomenon known as dissipative anomaly has been 
accorded the status of the ``zeroth-law" of turbulence \cite{KRS84,Zocchi,KRS98diss,Pearson,GotohVas,DB14}. 
A direct assessment of dissipation scaling is challenging because of the large time-scales of the quantities involved. % $\epsilon(\nu) L/{u^\prime}^3$
This means that both experiments and simulations require long run-times at ever-increasing Reynolds numbers.
%In contrast, the third-order inertial range moments has much  shorter time-scale with less stringent resolution requirements than higher-order moments - hence they can be measured with greater confidence. 

\vspace{0.2cm}
%In contrast, the third-order longitudinal absolute exponents can be measured with greater confidence hence they can be measured with greater confidence
%
In contrast, probing the validity of the zeroth law using the third-order longitudinal absolute scaling exponent is more favorable due to the following reasons.
Firstly, inertial range moments evolve over shorter time-scales than large-scale quantities which means that experiments and simulations require shorter run-times to capture their temporal evolution \cite{PK2020}. 
Secondly, third-order moments have less stringent resolution requirements than higher-order moments -- hence they can be measured with greater accuracy. Lastly, longitudinal velocity differences are one-dimensional cuts from the total velocity difference tensor and hence are more
feasible to measure in experiments. Despite these advantages, the third-order longitudinal absolute scaling exponent has largely been over-looked with some exceptions \cite{krs94,Gotoh02,krs98,dhruva2000,DBexp,krs22}, especially in the context of dissipation scaling. 
%%%%

\vspace{0.2cm}
Consequently, we have
highlighted the importance of the third-order longitudinal absolute exponent to the Reynolds number scaling of turbulent dissipation. Under the assumption of the Kolmogorov $4/5$ law we have shown that if $\xstar > 1$ then turbulent energy dissipation must vanish in the infinite Reynolds number limit, that is the zeroth law of turbulence will be violated. Alternatively, if the third-order longitudinal absolute exponent $ \xstar \le 1$, then dissipative anomaly can hold strictly.
%If $\xi_3 > 1$ then turbulent energy dissipation should vanish and kinetic energy will be conserved asymptotically
%with $\xi_3 -1$ serving as the bounding rate for the dissipation decay.
%%%%%%
%Furthermore, since the longitudinal velocity increment is known to scale linearly $S_3(r) \propto r$ in the inertial range \cite{krs98,KIKRS20}, it follows from Eq.~\ref{dissa.eq} that 
%$S_3(r) \to 0$ due to cancellations in its power-law prefactor. This implies that the velocity increment field will asymptotically become symmetric at least as far as its skewness is concerned. 
An examination of the third-order scaling exponents of the absolute velocity increments which appears to be a focal point in asserting the Reynolds number scaling of average dissipation will be reported as future work.

\section{Acknowledgments}
I am grateful to T. D. Drivas, G. L. Eyink, K. R. Sreenivasan and P.K. Yeung for helpful discussions.

\end{document}